\documentclass[graphicx,amsmath,amssymb,pra,twocolumn]{revtex4-1}

\usepackage{graphicx}		% Einbinden von Grafiken
\usepackage{color}
\usepackage{amsmath}
\usepackage{subfigure}
\usepackage{upgreek}

\usepackage{dsfont}

\begin{document}
	
	\title{Interplay of periodic dynamics and noise: insights from a simple adaptive system }	
	\author{Frederic Folz$^1$}
	\author{Kurt Mehlhorn$^2$}
	\author{Giovanna Morigi$^1$}
	\affiliation{$^1$Theoretische Physik, Universit\"at des Saarlandes, 66123 Saarbr\"ucken, Germany \\ $^2$Algorithms and Complexity Group, Max-Planck-Institut f\"ur Informatik, Saarland Informatics Campus, 66123 Saarbr\"ucken, Germany}
	
	\date{\today}
	
	\begin{abstract} 
We study the dynamics of a simple adaptive system in the presence of noise and periodic damping. The system is composed by two paths connecting a source and a sink, the dynamics is governed by equations that usually describe food search of the paradigmatic Physarum polycephalum. In this work we assume that the two paths undergo damping whose relative strength is periodically modulated in time and analyse the dynamics in the presence of stochastic forces simulating Gaussian noise.  We identify different responses depending on the modulation frequency and on the noise amplitude. At frequencies smaller than the mean dissipation rate, the system tends to switch to the path which minimizes dissipation. Synchronous switching occurs at an optimal noise amplitude which depends on the modulation frequency. This behaviour disappears at larger frequencies, where the dynamics can be described by the time-averaged equations. Here, we find metastable patterns that exhibit the features of noise-induced resonances.
\end{abstract}
	
	\maketitle
	
\section{Introduction}
Patterns are ubiquitous in nature, metastable spatio-temporal structures are observed from the microscopic to the astrophysical scale \cite{Cross:1993,Sherrington:2010,Zia:2011,Borgani:2012}. The systematic characterization of their onset and stability is a formidable challenge of theoretical physics. Theoretical descriptions are often based on coupled nonlinear equations for macroscopic variables, whose fixed points often capture essential features of the metastable dynamics, see for instance \cite{KPZ:1986,Cross:1993,Meron:2001,Steinbock:2019,Folz:2019}. 

A prominent example are the equations modelling the dynamics of biological systems, such as food search of Physarum polycephalum, a representative of the so-called true slime moulds \cite{Tero:2007}. Physarum polycephalum is  a single-celled organism that, despite its lack of any form of nervous system, is able to solve complex tasks like finding the shortest path through a maze \cite{Nakagaki:2000,Nakagaki:2004,Oettmeier:2020} and creating efficient and fault-tolerant networks \cite{Tero:2010,Boussard:2021}. These dynamics are qualitatively reproduced by the noise-free coupled nonlinear equations of motion \cite{Tero:2007}, which are a reference model system for optimization algorithms and deep learning \cite{Gao:2019}. 

Most theoretical descriptions of Physarum are noise-free and do not include the effect of a thermal bath in which Physarum is naturally immersed. On the other hand, tasks such as finding the optimal path in a maze are solved in contact with the external environment \cite{Nakagaki:2000,Tero:2010}, which shows that the Physarum dynamics is robust and probably even optimized for a certain level of noise, as typical of adaptive systems \cite{Gross:2005,Flies:2018}. 

Motivated by this question, in this work we consider the coupled nonlinear equations describing an adaptive system that can choose between two paths in the presence of Gaussian noise and that is inspired by the dynamics of Physarum polycephalum. We analyse the dynamics when the relative dissipation rate between the two paths is modulated in time and as a function of the modulation frequency and of the noise amplitude. We benchmark our results with the work of Ref.~\cite{Meyer:2017}, who studied this model for a fixed value of frequency and noise amplitude.   

This manuscript is organised as follows. In Sec.~\ref{Sec:Model} we introduce the physical model, in Sec.~\ref{Sec:Low:Frequency} we discuss the response as a function of the frequency and noise amplitude and identify the regime where stochastic resonance characterizes the dynamics. In Sec.~\ref{Sec:Large:Frequency} we then turn to the regime outside the stochastic resonance condition, when the frequency is sufficiently large and characterize the system dynamics as a function of the noise amplitude. The conclusions are drawn in Sec.~\ref{Sec:Conclusions}. The appendices provide details on the model and on the calculations in Secs. \ref{Sec:Low:Frequency}  and  \ref{Sec:Large:Frequency}.
	
\section{Two paths with modulated dissipation}
\subsection{The model}
\label{Sec:Model}

\begin{figure*}
		\includegraphics[width=1\textwidth]{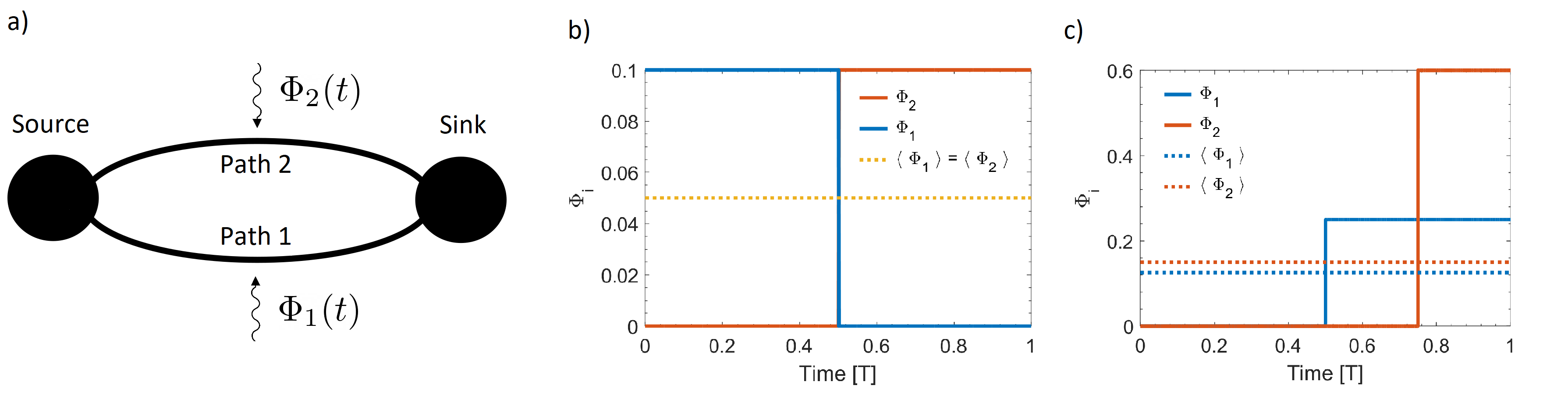}
		\caption{\label{Fig:1}(a) Illustration of a system consisting of two different paths connecting a source and a sink. The paths are exposed to a periodically modulated dissipation. Subplots (b) and (c) display the periodic function $\Phi_i(t)$, determining the  temporal behaviour of the damping rate of path $i$, over one period $T$ and for the two examples discussed in this work (symmetric (b) and biased case (c)). The time is in units of $1/\gamma$, the dotted lines are the time averages over one period.}
\end{figure*}

The model we consider is a simplified network, where a source and a sink are connected through two paths of equal and constant length $L$ as illustrated in Fig.~\ref{Fig:1}(a) and whose dynamics is inspired by Physarum polycephalum. Here, the capability to connect two food sources is modelled with the flow of gel inside the cell body along the network edges and is quantified by the conductivity $D_i$ of path $i=1,2$ \cite{Tero:2007,Bonifaci:2020}. This variable increases monotonically with the gel's flow and vanishes when the flow vanishes according to the deterministic equation ($i=1,2$):
\begin{equation}
\label{Eq:D:0}
[\partial_t D_i]_0=f(\tilde D_i)-\gamma_i(t)D_i\,,
\end{equation}
where $f(x)$ is the nonlinear force with argument $\tilde D_i=D_i/(D_1+D_2)$: 
\begin{eqnarray}
\label{Eq:f}
&&f(x) = \Gamma (1 + \epsilon) x^2 / (\epsilon + x^2)\,,
%&&\gamma_i(t) = \gamma + \Phi_i(t)\label{gamma}\,,
\end{eqnarray} 
with $\epsilon=0.2$ and $\gamma_i$ is the damping rate of path $i$ \cite{Meyer:2017}. The dissipation is here periodically modulated in time with period $T$, 
\begin{eqnarray}
\label{Eq:f}
%&&f(x) = (1 + \epsilon) x^2 / (\epsilon + x^2)\\
\gamma_i(t) = \gamma + \Phi_i(t)\label{gamma}\,,
\end{eqnarray}
with $\Phi_i(t)=\Phi_i(t+T)$ a periodic and positive function, see Fig.~\ref{Fig:1}. Let $\omega = 2 \pi/T$ denote the angular frequency. Furthermore, we set $\Gamma = \gamma$ in the following. From now on, the time will be given in units of $\gamma^{-1}$ unless otherwise stated.

In this work we characterize the system response to the dynamically changing environment in the presence of the stochatic force $\xi_i(t)$:
\begin{equation}
\label{Eq:D}
\partial_t D_i=[\partial_t D_i]_0+\tilde\alpha\xi_i(t)\,.
\end{equation}
The stochastic force is scaled by the positive parameter $\tilde \alpha = \alpha \gamma$ and describes white noise. It has vanishing expectation value, $\langle \xi_i(t)\rangle = 0$, and correlations
\begin{align}
\label{Eq:xi}
%    \langle \xi_i \rangle & = 0 \\
    \langle \xi_i(t) \xi_j(t') \rangle & = \kappa \delta_{i, j} \delta(t - t') \,,
\end{align}
where $\kappa = \gamma^{-1}$ and $\langle \cdot \rangle$ is the average taken over a sufficiently large number of independent realizations of the stochastic process $\xi_i(t)$ \cite{Haken:1983,vanKampen} (Note that, since this force can take negative value, we need to regularize the behaviour of Eq.~\eqref{Eq:D} for small values of $D_i$ so as to keep $D_i\ge 0$. This in turn reflects the physical constraint that $D_i$ represents a conductivity. For this purpose, when $D_i = 0$ and the right-hand-side of Eq.~\eqref{Eq:D} becomes negative, we set it equal to 0.) In what follows we numerically simulate Eq.~\eqref{Eq:D} using the Euler-Maruyama scheme \cite{Kloeden} with a step size $\Delta t = 0.05 \gamma^{-1}$, unless otherwise specified.

As in Ref.~\cite{Meyer:2017} we quantify the system's response by means of the quantity
\begin{align}
        \label{Eq:c}
	    c(t) = \frac{D_1(t) - D_2(t)}{D_1(t) + D_2(t)}\,,
\end{align}
which we denote by risk function. The risk function varies in the interval $[-1,1]$. The extremal values $c=+1$ and $c=-1$ correspond to the system being in the path $i=1$ and $i=2$, respectively. We note that the function minimizing the risk takes the form
$$c_r(t) =2\left( \theta(\Phi_2(t) - \Phi_1(t))-\frac{1}{2}\right)\,,$$
with $\theta(x)$ Heaviside's function. The corresponding dynamics follows the path whose instantaneous dissipation is minimal. 

\subsection{About the biological system.}

Eq. (1) has been proposed in Ref.\ \cite{Tero:2007} for simulating the dynamics of Physarum polycephalum for a constant dissipation $\gamma_i$. This model was extended in Ref. [18] for the purpose of analysing the effect of noise on the adaptivity of Physarum polycephalum to dynamically varying environmental conditions. The dynamically varying environment was there modelled by a periodically modulated dissipation. The latter simulated the effect of light, which inhibits the growth of Physarum polycephalum. \\
\indent In the present work we start from that analysis and extend it by investigating the dynamics as a function of the noise strength and of the modulation frequency. This allows us to identify whether and when there is an optimal noise strength for which the system optimally adapts to the external environment.

\section{Symmetric configuration}
\label{Sec:Low:Frequency}

In the following we determine the dynamics as a function of $\alpha$, the noise strength, and $\omega$, the frequency at which dissipation is modulated. 
\begin{figure*}
    \centering
    \includegraphics[width=1\textwidth]{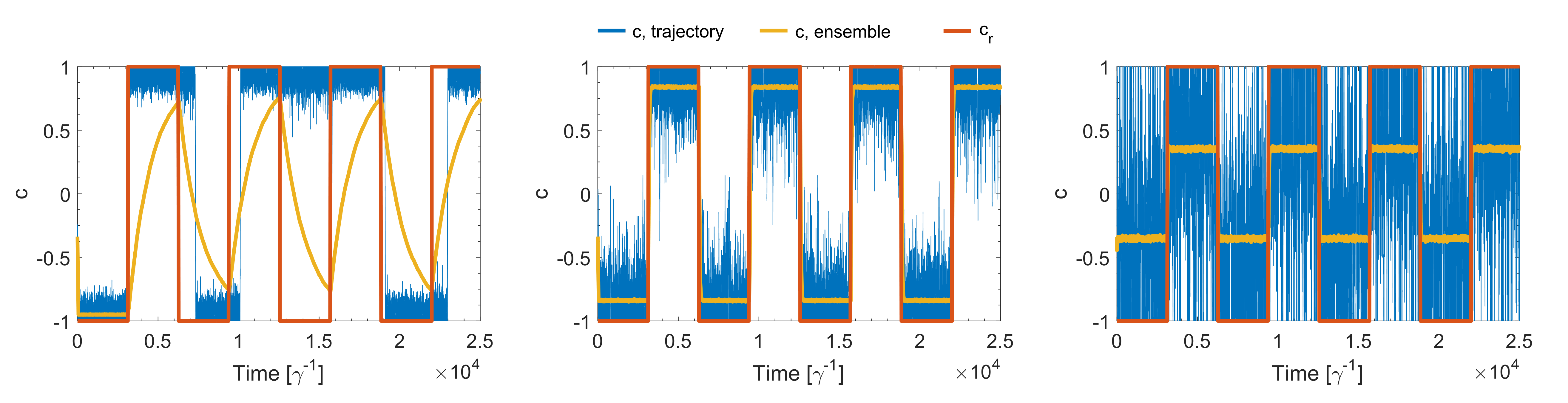}
    \caption{\label{fig:Stochastic}(color online) Time evolution of the risk function $c$ for $\omega = 10^{-3} \gamma$ and (a) $\alpha= 0.051$, (b) $\alpha = 0.099$ and (c) $\alpha = 0.222$. The blue line corresponds to one trajectory, the yellow line is the average over 5000 trajectories, the red line displays $c_r(t)$. The initial conditions are $D_1^0 = 0.5$, $D_2^0 = 1$ corresponding to $c^0 = -1/3$. In Appendix \ref{App:A} we report a zoom of subplot (a).}
\end{figure*}

We assume that the two functions $\Phi_i(t)$ are step functions shifted by half period with respect to one another, namely, $\Phi_1(t)=\Phi_2(t+T/2)$. Over one period we choose $\Phi_1(t)=\gamma_0\,\theta(t)\theta(T/2-t)$, with $\gamma_0=0.1\gamma$. Figure \ref{fig:Stochastic} displays the evolution of the conductivity $D_1$ for different, increasing values of $\alpha$. Among the three examples displayed, the flow seems to best adapt to the periodic changes of the external parameters for $\alpha\sim 0.1$ (we emphasize that for $\alpha=0$ the system does not switch path).

We quantify the capability of the system to adapt to the changes of dissipation over the evolution time $[0,t_{\rm end}]$ by means of the normalized correlation function
\begin{align}
    \label{Eq:g}
    g(\tau) = \frac{2}{t_{\rm end}} \int_0^{t_{\rm end}} \left(\theta(c(t))-\frac{1}{2}\right)c_r(t - \tau)  dt
\end{align}
which quantifies the overlap between the signal $c(t)$ and the function $c_r(t)$, as a function of the delay $\tau>0$. Perfect correlation (anticorrelation) between dynamics and dissipation minimizes (maximizes) the risk and corresponds to $g = 1$ and $\tau \to 0$ ($\tau=T/2$). 

	\begin{figure}
		\includegraphics[width=0.45\textwidth]{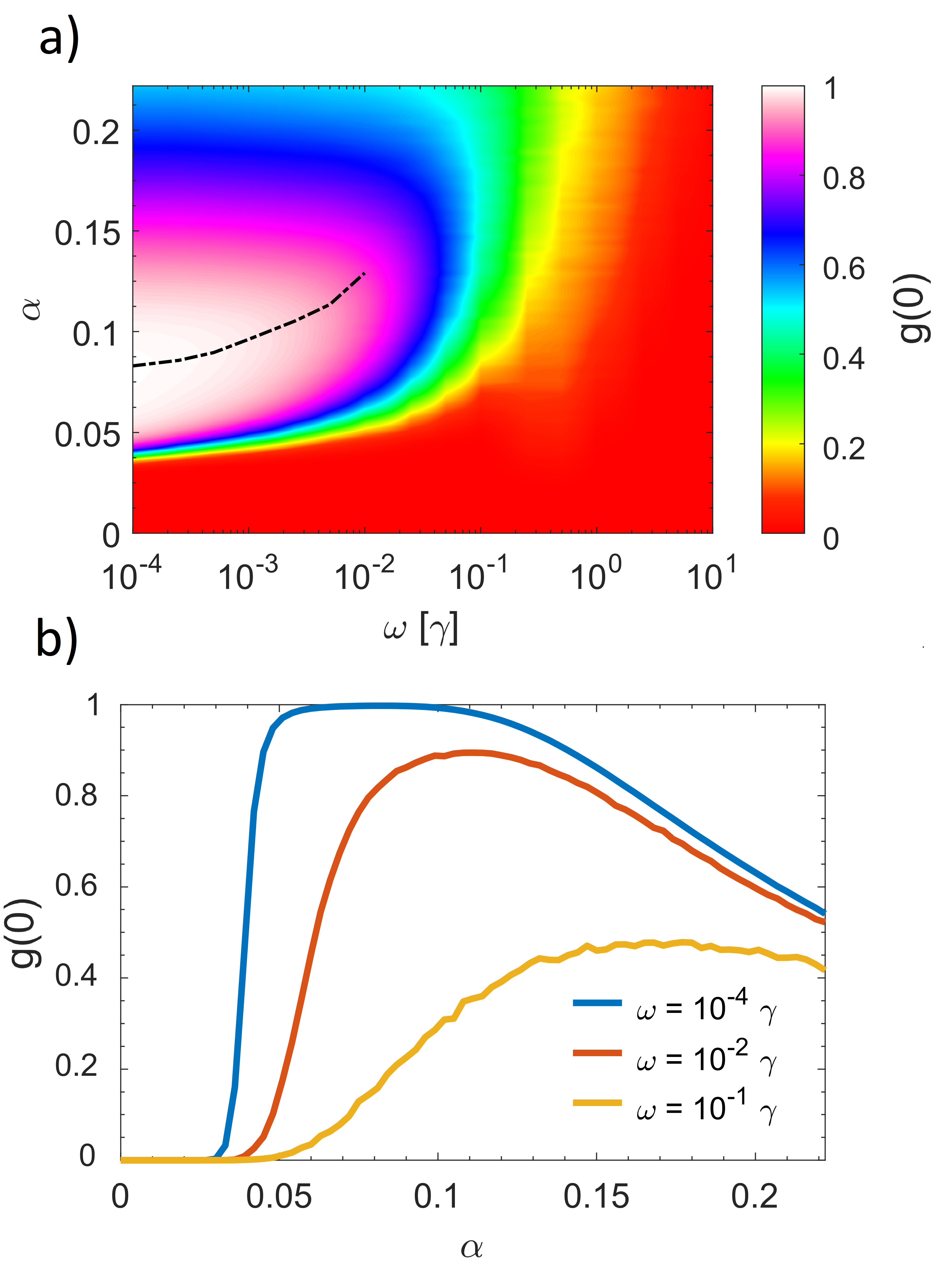}
		\caption{\label{fig:Probability}(color online) (a) Color plot of the measure $g(0)$, caclulated using Eq.~\eqref{Eq:g}, as a function of the noise strength $\alpha$ and of the modulation frequency $\omega$. The dotted line represents the noise strengths at which we expect stochastic resonance to occur according to the matching condition of Eq.~\eqref{Eq:resonance}. For the displayed range of angular frequencies $\omega$ it holds $g(0) \approx \max_{\tau}(g(\tau))$. (b) Measure $g(0)$ as a function of $\alpha$ for three values of the frequency $\omega$, see legenda. We used the initial conditions $D_1^0 = 0.5$, $D_2^0 = 1$ corresponding to $c^0 = -1/3$. The data was generated by evaluating an ensemble of 5000 trajectories with simulation time of $t_{\rm end} = 8 \pi /\omega$ and step size $\Delta t = 10^{-4} /\omega$. }
	\end{figure}	
	
Figure \ref{fig:Probability}(a) displays the colour plot of $g(0)$ as a function of the noise strength $\alpha$ and of the modulation frequency $\omega$. For the full range of angular frequencies $\omega$ shown in the figure, it holds $g(0) \approx \max_{\tau}(g(\tau))$. We observe a region for which $g(0)\approx 1$, indicating synchronous behaviour at an optimal noise strength within a frequency range $\omega \lesssim 10^{-2}\gamma$. In subplot (b) we display $ g(\tau)$ for three values of $\omega$  and as a function of $\alpha$: The dynamics exhibits the features of stochastic resonance \cite{Gammaitoni:1998}, the optimal noise strength where $g(\tau)$ is maximum can be found from the relation 
\begin{align}
        \label{Eq:resonance}
	    t_s=\frac{\pi}{\omega} 
\end{align}
where  $t_s$ is the average switching time between the two stable fixed points $c=-1$ and $c=1$. We estimate the switching time using a one-dimensional model, the details are reported in Appendix \ref{App:B}. The resulting curve is the dotted line of Figure \ref{fig:Probability}(a) and reproduces the position of the resonance in the $\alpha-\omega$ plane which we find numerically. 

The resonance behaviour as a function of $\alpha$ broadens as $\omega$ increases. For $\omega\gtrsim\gamma$ there seems to be no correlation between $c(t)$ and the temporal modulation of the dissipation. In the next section we analyse the effect of noise in this regime.

\section{Secular regime}
\label{Sec:Large:Frequency}

 In this section we analyse the dynamics as a function of the noise strength at large frequencies, such that $\omega/\gamma\gg 1$. In this regime we expect that the effect of the time-dependent dissipation on the dynamics can be replaced by its average. We choose an asymmetric modulation of dissipation between the two paths and fix $\omega = 10\gamma$. In order to compare with Ref.~\cite{Meyer:2017} we define
 \begin{eqnarray}
&&\Phi_1(t)=\gamma_{01}\theta(t-T/2)\theta(T-t)\\
&&\Phi_2(t)=\gamma_{02}\theta(t-3T/4)\theta(T-t)
\end{eqnarray} for $t\in[0,T]$, with $\gamma_{01}=\gamma/4$ and $\gamma_{02}=3\gamma/5$. According to this choice we expect a bias towards path 1, since it is characterized by the minimal (time-averaged) dissipation. We note that in Ref.~\cite{Meyer:2017} the authors considered the specific noise strength $\alpha = 0.05$ and simulated the dynamics till the time $t_{\rm end} = 200 \pi \gamma^{-1}$. In the following we study the dynamics as a function of $\alpha$. Moreover, we analyse the convergence of the simulated trajectories by taking different times and by comparing the results with the stationary state, which we analytically determine.

\begin{figure}
		\includegraphics[width=0.45\textwidth]{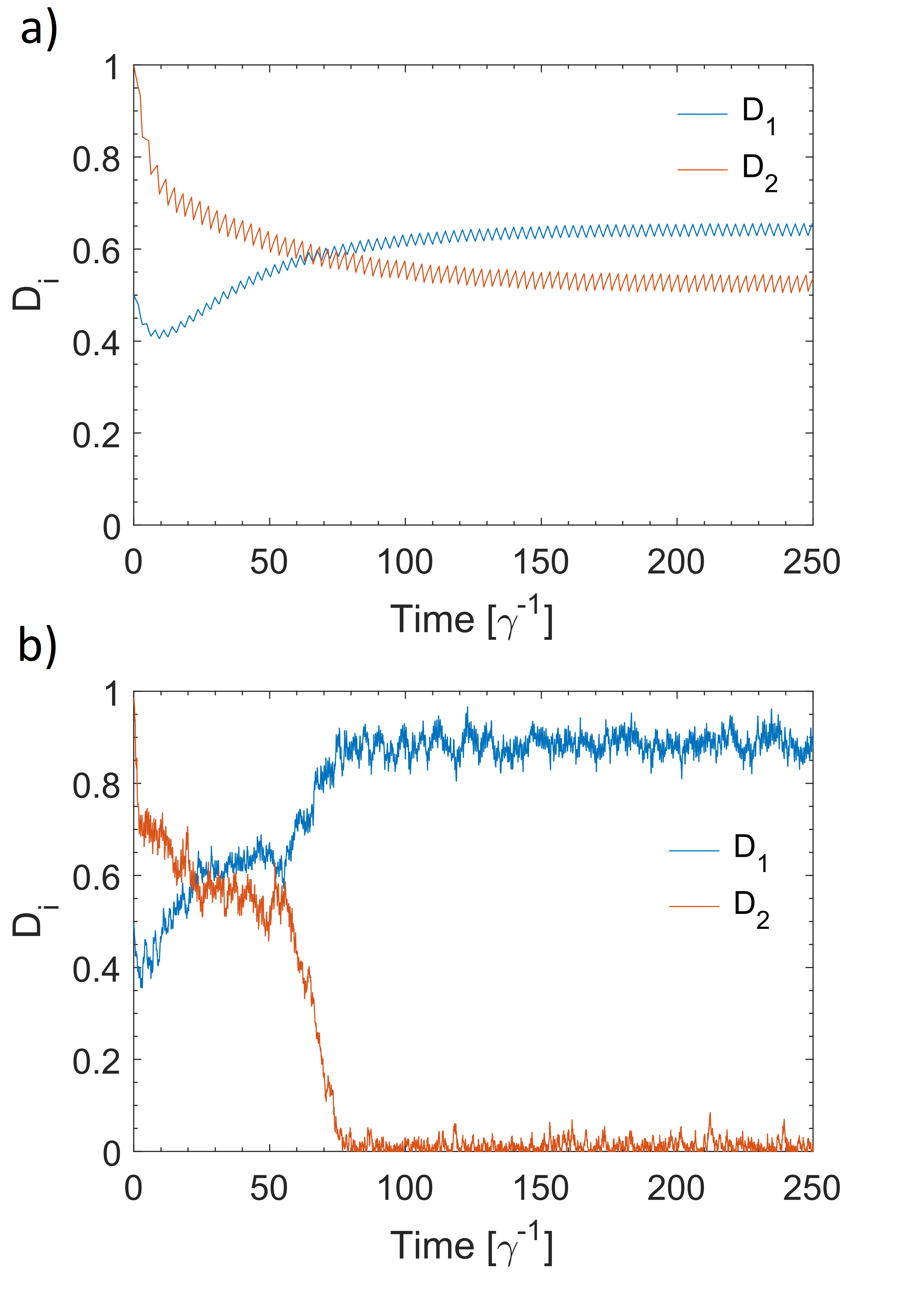}
		\caption{\label{fig:2}(color online) Evolution of the conductivities $D_i$ as a function of time (in units of $1/\gamma$) for $\omega=10\gamma$ and (a) $\alpha = 0$, (b) $\alpha = 0.05$. In the deterministic case, $\alpha=0$, the system converges to the fixed point of the secular equation, see the blue solid circle of Fig.~\ref{fig:3}(a). In the presence of noise this fixed point is metastable: the system quickly converges to the stable fixed point at which the system favors the path exposed to minimal averaged dissipation. The data shown represents a single trajectory. The initial conditions are $D_1^0 = 0.5$ and $D_2^0 = 1$.}
	\end{figure}

Figure \ref{fig:2} displays the evolution of the conductivities $D_i$ by numerically integrating Eq.~\eqref{Eq:D} assuming that initially $D_2>D_1$. In the deterministic case, for $\alpha=0$, the curves exhibit fast, small-amplitude oscillations at a frequency close to the modulation frequency $\omega$, their time-average over a period varies over a significantly longer time scale. After a transient, the conductivity $D_1$ reaches a value larger than $D_2$. The corresponding dynamics in the presence of white noise is shown in subplot (b): We observe a time scale separation as in the deterministic case, while the frequency of the fast oscillations become 
chaotic. The slow dynamics reaches a metastable state (corresponding to the asymptotic state of the noise-free dynamics) where it is trapped for a relatively short time. It then quickly reaches the steady state, with the value of the conductivity $D_1$ approaching unity, while the flow along the second path is almost suppressed.

\subsection{Fixed points of the secular dynamics}

In order to perform an analytical study, we consider the secular dynamics, where we average the equations Eq.~\eqref{Eq:D} over a period $T$ of the oscillations and replace $\gamma_i(t)$ with the time-averaged dissipation coefficients:
$$\gamma_i^{\rm eff}=\gamma+\langle \Phi_i\rangle_T\,,$$
with 
$$\langle A\rangle_T \equiv \frac{1}{T}\int_t^{t+T}{\rm d}\tau A(\tau)\,.$$
According to our parameter choice $\gamma_1^{\rm eff}<\gamma_2^{\rm eff}$ and in the regime of validity of this secular approximation path 1 is favoured. We then set $\alpha=0$ and study the fixed points of the dynamics $D_i^*$, fulfilling $\partial_t D_i^*=0$.  We report the details in Appendix \ref{App:C}. 

Figure \ref{fig:3}(a) displays the fixed points $(D_1^*, D_2^*)$: stable (unstable) solutions are represented by solid (hollow) circles. The vector fields illustrate the flow.  The path with average minimal dissipation corresponds to the green circle. The blue circle is the metastable path to which the deterministic dynamics of Fig.~\ref{fig:2}(a) converges for the given initial condition. This solution still characterizes the transient dynamics in the presence of noise, Fig.~\ref{fig:2}(b). On longer time scales, however, the stochastic dynamics brings the system to the path minimizing dissipation.

\begin{figure}
		\includegraphics[width=0.45\textwidth]{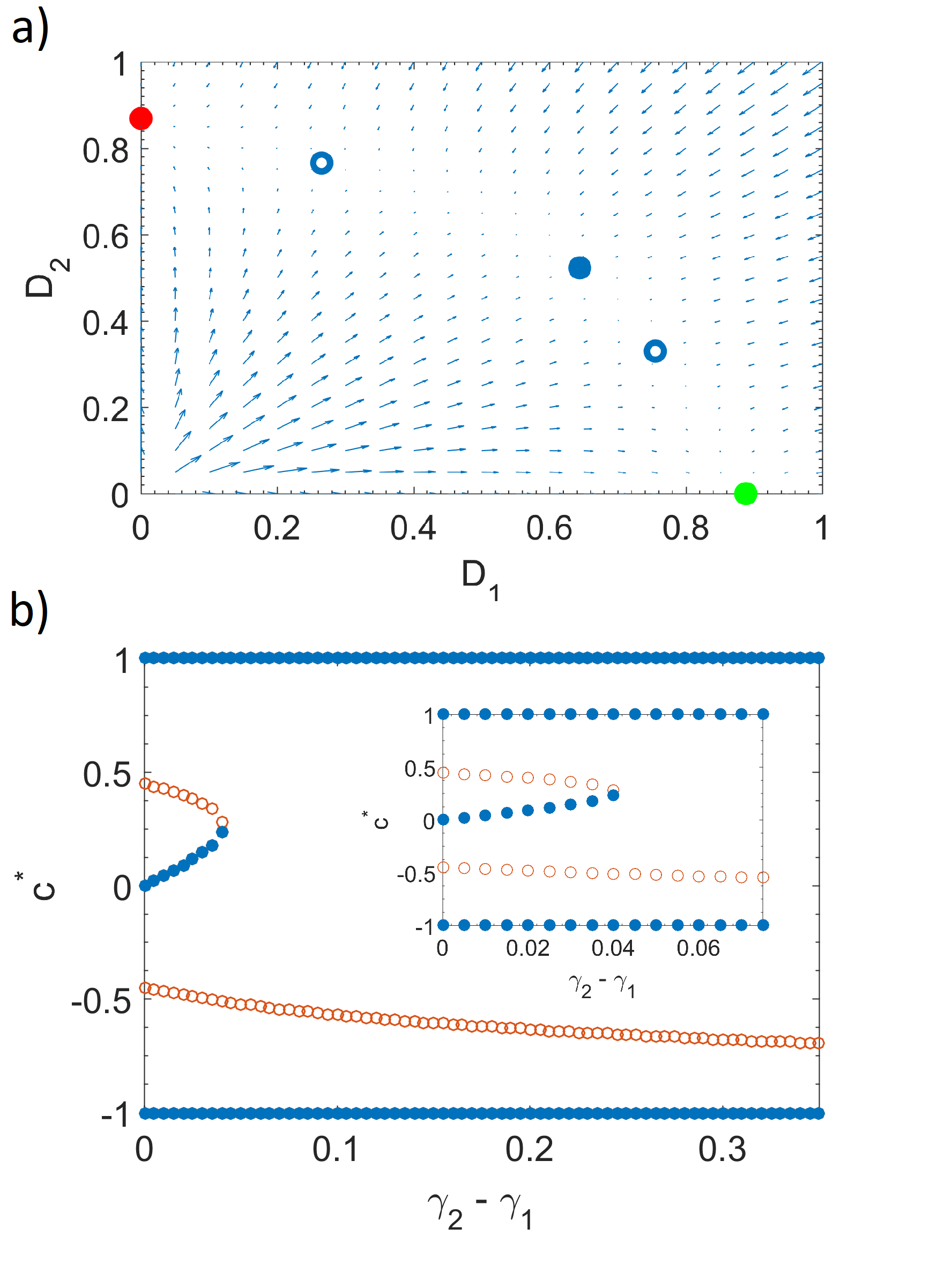}
		\caption{\label{fig:3}(color online) (a) Fixed points $(D_1^*, D_2^*)$ of the secular equations  for $\langle \Phi_1 \rangle_T = 0.125$ and $\langle \Phi_2 \rangle_T = 0.15$. The arrows indicate the flow. Stable (unstable) fixed points are represented by solid (hollow) circles. The green (red) stable fixed point correspond to the system choosing the path with minimal (maximal) average dissipation. (b) Fixed points $c^* = (D_1^* - D_2^*)/(D_1^* + D_2^*)$ as a function of $\gamma_2^{\text{eff}} - \gamma_1^{\text{eff}}$ with fixed $\gamma_1^{\text{eff}} = 1.125$. Full (hollow) circles indicate stable (unstable) fixed points. The inset zooms into the parameter region of the intermediate, metastable fixed point.}
	\end{figure}
Figure \ref{fig:3}(b) shows the variable
 $c^* = (D_1^* - D_2^*)/(D_1^* + D_2^*)$
as a function of the average dissipation $\gamma_2^{\rm eff}$, which is varied starting from the symmetric case $\gamma_2^{\rm eff}=\gamma_1^{\rm eff}$ while keeping $\gamma_1^{\rm eff}$ fixed. The blue solid (red hollow) symbols indicate the stable (unstable) solutions. The red hollow symbols are the borders of the basins of attraction of the nearby stable solution for the deterministic dynamics. The two extremal solutions - where the system settles on one of the two paths - are independent of $\gamma_2^{\rm eff}$. Instead, the solution characterized by finite conductivities along both paths exists solely below a threshold value $\gamma_{2,th}^{\rm eff}$. As $\gamma_2^{\rm eff}$ increases from $\gamma_1^{\rm eff}$\ towards the threshold, this fixed point moves towards positive values indicating that the symmetry of the solutions is broken.  Correspondingly, its basin of attraction becomes asymmetric by increasing $\gamma_2^{\rm eff}$ and biased towards the region $c>0$.  At the threshold, the basin of attraction of  $c^*=1$ undergoes a discontinuous jump and extends to the neighbourhood of the other fixed point $c^*=-1$. 

Despite the fact that this analysis has been performed using a secular approximation, we note that the fixed points of \ref{fig:3}(a) allow us to understand the steady state of Fig.~\ref{fig:2} as well as the metastable dynamics in  Fig.~\ref{fig:2}(b). According to this picture, one would expect that by increasing $\alpha$ the rate of convergence to the stable state shall accordingly increase. In the following section we will show that this is only partly true.

\subsection{Convergence to the stationary state}

	\begin{figure}
		\includegraphics[width=0.45\textwidth]{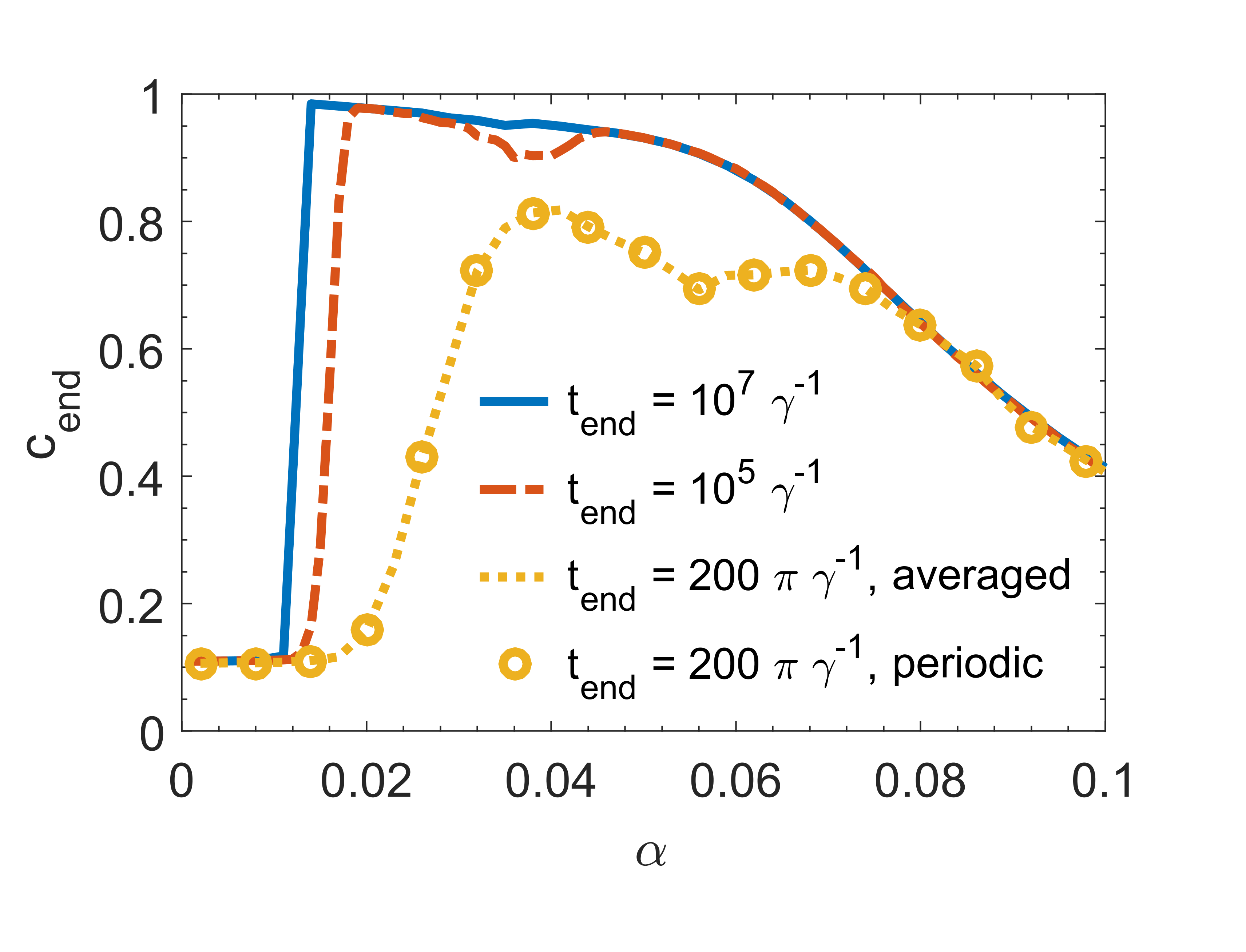}
		\caption{\label{fig:cend_0}(color online) Average value $\bar c$ at the simulation time $t_{\rm end}$ as a function of the noise strength $\alpha$ for $t_{\rm end}= 200 \pi \gamma^{-1}, 10^5 \gamma^{-1},10^7 \gamma^{-1}$. The data shown was generated by averaging over a total of 5000 trajectories. The initial conditions are $D_1^0 = 0.5$, $D_2^0 = 1$ corresponding to $c^0 = -1/3$. For all values of $t_{\rm end}$ we evolve the secular equation. We also show the result of the full, periodic dynamics (circles) for $t_{\rm end}= 200 \pi \gamma^{-1}$. Here, in order to generate comparable data, we use the same set of random numbers for the trajectories.}
	\end{figure}
	
The analysis of the fixed point of the deterministic dynamics provides some insight into the behaviour observed in Fig.~\ref{fig:2}: Here, metastable fixed points corrrespond to metastable configurations whose lifetime is limited by noise. Fluctuations, in general, allow the system to explore a large configuration space. 

In order to characterize the convergence to the steady state as a function of the noise strength $\alpha$ we first determine the variable $c$ for sufficiently large integration times $t_{\rm end}$. We single out the slowly-varying value by averaging out the fast oscillations over a period,
\begin{align}
        \label{Eq:ctend}
	    \bar c(t_{\rm end}) = \frac{1}{T} \int_{t_{\rm end} - T}^{t_{\rm end}} c(t') dt'\,,
\end{align}
where $c(t)$ in the integrand is the ensemble average over the individual trajectories. Figure \ref{fig:cend_0} displays $\bar c(t_{\rm end})$ as a function of the noise strength $\alpha$ for different integration times $t_{\rm end}$. The choice $t_{\rm end}=200\pi/\gamma$ corresponds to the same integration time of Ref.~\cite{Meyer:2017}. Comparison with longer simulation times shows that at this time and $\alpha\lesssim 0.07$ the dynamics has not yet converged to the stationary state. For $\alpha\lesssim 0.02$, moreover, the system is still trapped in the metastable configuration at $c^*\sim 0$ even for the longest integration time here considered. 

\begin{figure*}
		\includegraphics[width=1\textwidth]{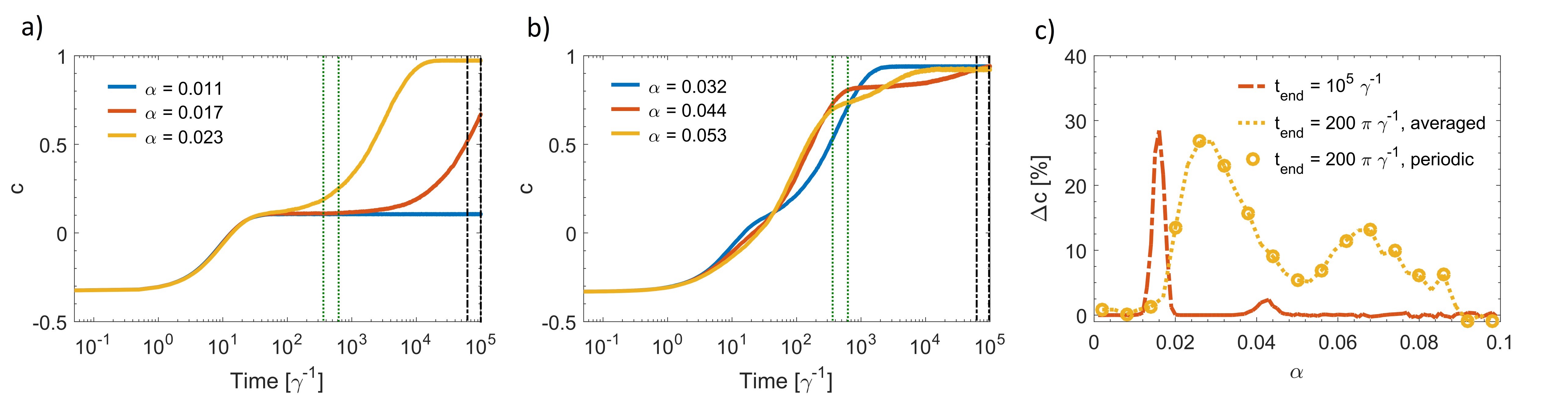}
		\caption{\label{fig:cend}(color online) Time evolution of the average value of $c$, Eq.~\eqref{Eq:c}, for the noise strengths (a) $\alpha = 0.011$, $\alpha = 0.017$, $\alpha = 0.023$ and (b) $\alpha = 0.032$, $\alpha = 0.044$, $\alpha = 0.053$. In each subplot the dotted green (black) lines label the times $0.6\cdot 200 \pi \gamma^{-1}$ ($0.6 \cdot 10^5/\gamma$) and $t = 200 \pi \gamma^{-1}$ ($10^5 \gamma^{-1}$). (c) Variation $\Delta c$ as a function of $\alpha$ for the simulation times $t_{\text{end}} = 200 \pi \gamma^{-1}$ and $t_{\text{end}} = 10^5 \gamma^{-1}$. For $t_{\rm end} = 200 \pi \gamma^{-1}$ the trajectories for periodic and time-averaged dynamics are shown. To generate comparable data, in both of these cases the same set of random numbers was used to simulate the trajectories. The integration method, trajectories, and initial conditions are the same as in Fig.~\ref{fig:cend_0}.}
\end{figure*}

The behaviour in the interval $0.02\lesssim\alpha\lesssim 0.07$ is remarkable, as it exhibits local minima which have the form of resonances. We note that the lifetime of the metastable configurations at $\alpha\sim 0.04$ exceeds $t_{\rm end}\sim 10^5/\gamma$. 
Figure \ref{fig:cend} shows the time evolution of $c$ about these special values of $\alpha$. Subplot (a) displays the time evolution for values of $\alpha\lesssim 0.02$. Here, the system is trapped in the metastable configuration corresponding to the fixed point at $c^*\gtrsim 0$: the residence time visibly decreases as $\alpha$ increases. This trend is still visible at larger values of $\alpha$ in subplot (b). At longer time scales, however, the curves in (b) show a metastable regime close to path 1 where the system remains trapped and whose lifetime is maximal for $\alpha\sim 0.04$. This metastable configuration is not captured by the fixed point analysis and seems to crucially depend on the noise strength. It has thus the form of a noise-induced resonance. 
We study these resonances by inspecting the variation $\Delta c$  of the curve $c(t)$ at the extremal of the interval of time $\mathcal I(\delta)=[\delta \cdot t_{\rm end}, t_{\rm end}]$ with $\delta \in (0, 1)$:
	\begin{align}
	    \Delta c = \Big (c_{\rm end} - \frac{1}{T} \int_{\delta\cdot t_{\rm end} - T}^{\delta\cdot t_{\rm end}} c(t') dt' \Big ) c_{\rm end}^{-1}\,,
	\end{align}
where $c_{\rm end}$ is the stationary value for $t\to\infty$. When the system dynamics does not change over the interval $\mathcal I(\delta)$ then the variation $\Delta c=0$. Figure \ref{fig:cend}(c) shows $\Delta c$ as a function of $\alpha$ for $\delta=0.6$ and different integration times $t_{\rm end}$. Metastable configurations appear as resonances. We observe several resonances for relatively short integration times, while for increasing $t_{\rm end}$ the number of metastable states decreases.  The small resonance at $\alpha\sim 0.04$ signals the metastable configuration of Fig.~\ref{fig:cend}(b). The largest resonance at $\alpha \sim 0.02$ separates the regime where the system is still trapped in the metastable fixed point from the regime where the system has already escaped this region.

\begin{figure*}
\includegraphics[width=1\textwidth]{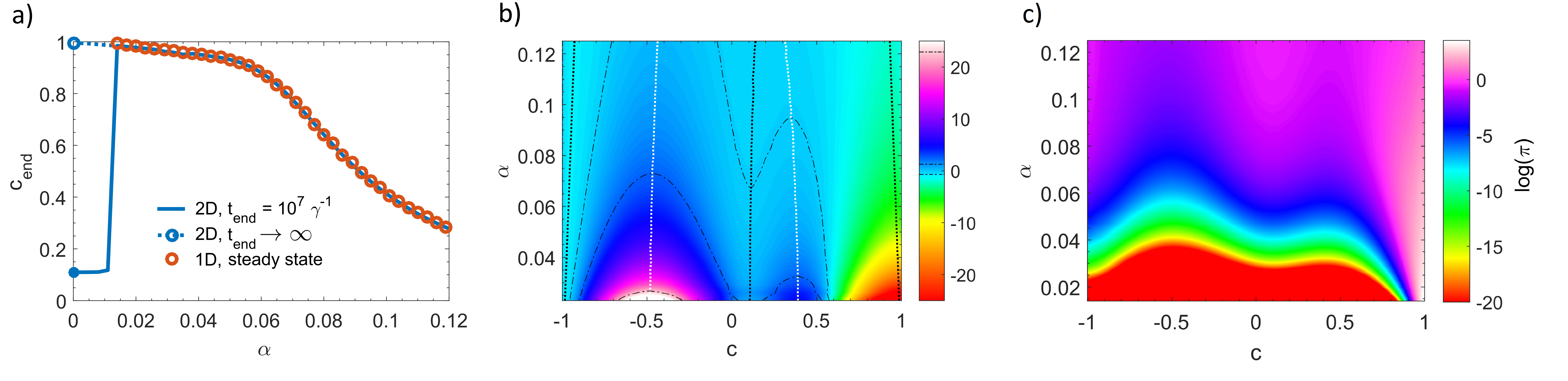}
		\caption{\label{fig:1D-2D}(color online) (a) The risk function $c_{\rm end}$ as a function of the noise strength $\alpha$ for the parameters of Fig.~\ref{fig:cend_0}. The prediction of the model of Eq.~\eqref{Eq:D} at $t_{\rm end} = 10^7 \gamma^{-1}$ is compared with the one of the Fokker-Planck equation of Eq.~\eqref{Eq:FPE} for $t\to\infty$. For the Fokker-Planck approach, the value of $c_{\rm end}$ is given by the average of the steady state distribution $c_{\rm end} = \int_{-1}^1 \tilde c p_0(\tilde c) d\tilde c$. Furthermore, we only take $\alpha > 0.01$ for the one-dimensional case as for smaller noise strengths the numerical algorithm does not converge within the considered simulation time. (b) Color plot of the potential $\varphi$, Eq.~\eqref{Eq:Phi}. The dotted white lines indicate the local maxima, the dotted black lines the local minima. (c) Color plot of the logarithm of the probability density distribution $p_0$, Eq.~\eqref{Eq:stationary}, as a function of $\alpha$ and $\tilde c$.}
	\end{figure*}

\subsection{Steady state}
\label{Sec:FPE}

We complete this study by discussing the steady state of the dynamics as a function of $\alpha$. We apply the approach implemented in Ref.~\cite{Meyer:2017}, and extend it to determine the dependence on $\alpha$. We review here the basic steps. The approach consists in determining the time averaged dynamics of the single variable $\tilde c$, assuming that it undergoes a time-continuous Markov process in the presence of a drift $\mu(\tilde c)$ and an Ito-diffusion with amplitude $\sigma(\tilde c,t)$, which are determined by means of an equation-free analysis, see Appendix \ref{App:D}. We verify the validity of this approximation by comparing the predictions of the full dynamics at $t_{\rm end}=10^7/\gamma$  with the one of the stochastic equation for the single-variable $\tilde{c}$, see Fig.~\ref{fig:1D-2D}(a). For the one-dimensional case we take noise strengths $\alpha > 0.01$ since for smaller values the numerical algorithm does not converge within the simulation times we considered. The corresponding Fokker-Planck equation for the probability distribution $p(\tilde c, t)$ takes the form 
\begin{align}
\label{Eq:FPE}
	    \partial_t p (\tilde c, t) = - \frac{\partial}{\partial \tilde c} \mathcal J(\tilde c,t)\,,
	    \end{align}
with the current:
\begin{equation} 
\label{Eq:J}
 \mathcal J(\tilde c,t)=\mu(\tilde c)p(\tilde c, t)-\frac{1}{2}\frac{\partial}{\partial \tilde c} \sigma^2(\tilde c)p(\tilde c,t)\,.
\end{equation}
We denote the steady state of $p(\tilde c, t)$ by $p_0(\tilde c)$. The steady state distribution $p_0(\tilde c)$ fulfills $\partial_tp_0(\tilde c)=0$ which corresponds to a constant current $\mathcal J$ and explicitly reads:  
\begin{align}
        \label{Eq:stationary}
	    p_0(\tilde c) = \frac{\mathcal N}{\sigma^2(\tilde c)}\exp\left(- \varphi(\tilde c)\right)
\end{align}
with $\mathcal N$ a normalization coefficient, warranting $\int_{-1}^1d\tilde c \, p(\tilde c)=1$, and 
\begin{align}
\label{Eq:Phi}
	    \varphi(\tilde c) =-\int_{-1}^{\tilde c} \frac{2 \mu(y)}{\sigma^2(y)} dy
\end{align}
being the potential associated with the stationary solution. Potential and steady state probability are shown in Fig.~\ref{fig:1D-2D}(b)  and (c), respectively, as a function of $\tilde c$ and $\alpha$. For small values of $\alpha$ the minima and maxima are localized at the fixed points of the deterministic equation. The position of the minima are shifted towards the center of the interval as $\alpha$ is increased. Increasing the noise, moreover, decreases the barrier between minima: at sufficiently large $\alpha$ the system explores the full interval of values of $\tilde c$ with longer residence times in the region at $\tilde c=1$. At large $\alpha$ the effect of noise is to diffuse the solution across both paths keeping a bias towards $c=1$. We remark that the noise-induced resonances observed in Fig.\ \ref{fig:cend_0} are not captured by the equilibrium potential calculated from the one-dimensional Fokker-Planck equation.
		
\section{Conclusion}
\label{Sec:Conclusions}

We have analysed the dynamics of a simple adaptive system as a function of the strength of a stochastic force. The system is composed by two paths connecting a sink and a source and subject to a periodic modulation of the dissipation rate between the two paths. 

When the dissipation modulation frequency is smaller than the mean value of the dissipation rate, the system dynamics exhibit stochastic resonance, with the system periodically switching to the path minimizing dissipation. At large frequencies, instead, the dynamics is reproduced by the secular equations, obtained by taking the time-average of the damping coefficient over a period. The steady state is the fixed point of the secular equations which minimizes dissipation. The metastable configurations are in general the other fixed points of the secular equations, and the net effect of noise is to limit their life-time. Nevertheless, the dynamics also exhibits other metastable configurations at certain values of the noise amplitude that are neither captured by stochastic resonance nor by the fixed point analysis. They exhibit the features of noise-induced resonances.  

Our study suggests that noise could play a non-trivial role in both developing and optimizing algorithms for search problems, network design and artificial intelligence. In the future we will extend this investigation to a network such as the configuration considered in Ref.~\cite{Bonifaci:2020} in the presence of a dynamically-changing environment.

\acknowledgements
    
The authors acknowledge support from the  Deutsche  Forschungsgemeinschaft (DFG, German Research Foundation) Project-ID No.429529648, TRR 306 QuCoLiMa (Quantum Cooperativity of Light and Matter). 
	
\begin{appendix}

\section{Time evolution of the risk function}
\label{App:A}

\begin{figure}
\includegraphics[width=0.45\textwidth]{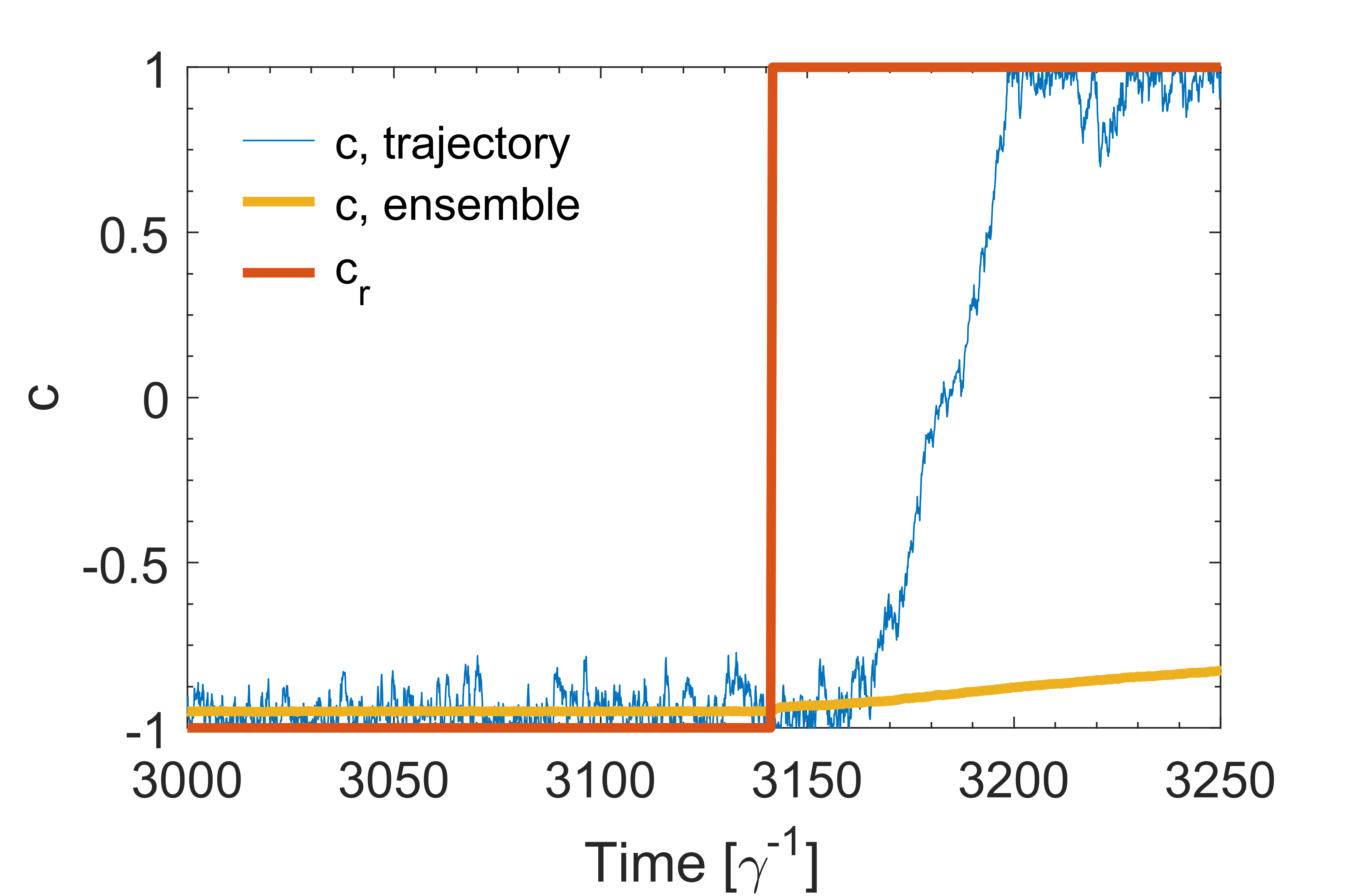}
		\caption{\label{fig:StochasticInset}(color online) Time evolution of the risk function $c$ for $\omega = 10^{-3} \gamma$ and $\alpha= 0.051$. The blue line corresponds to one trajectory, the yellow line is the average over 5000 trajectories, the red line displays $c_r(t)$. See Fig.\ \ref{fig:Stochastic}(a).}
	\end{figure}

Figure \ref{fig:Stochastic}(a) shows a single trajectory of the time evolution of the risk function $c$ for $\omega = 10^{-3} \gamma$ and $\alpha = 0.051$. Figure \ref{fig:StochasticInset} zooms over the behaviour during one period.

\section{Switching time}
\label{App:B}

We determine $t_s(\alpha,\omega)$, Eq.~\eqref{Eq:resonance}, using the first-passage time. This is the time to reach a metastable point, say, $c =-1$, when starting close to the other metastable point, say, $c = 1-\epsilon$, in an interval with reflecting boundaries. This corresponds to the first passage time of the one-dimensional model of Sec.~\ref{Sec:FPE} and takes the form \cite{vanKampen,Meyer:2017b}
\begin{align}
\label{Eq:first-passage}
	    t_s(\alpha,\epsilon) = \int_{1-\epsilon}^{-1}dy\frac{2}{\sigma^2(y) p(y)} \int_1^y dz p(z) .
\end{align}
We evaluate $ t_s(\alpha,\epsilon)$ numerically and use it in Eq.~\eqref{Eq:resonance} in order to find the stochastic resonance condition.
	
\section{Fixed points and linear stability analysis}
\label{App:C}

\begin{figure}
		\includegraphics[width=0.45\textwidth]{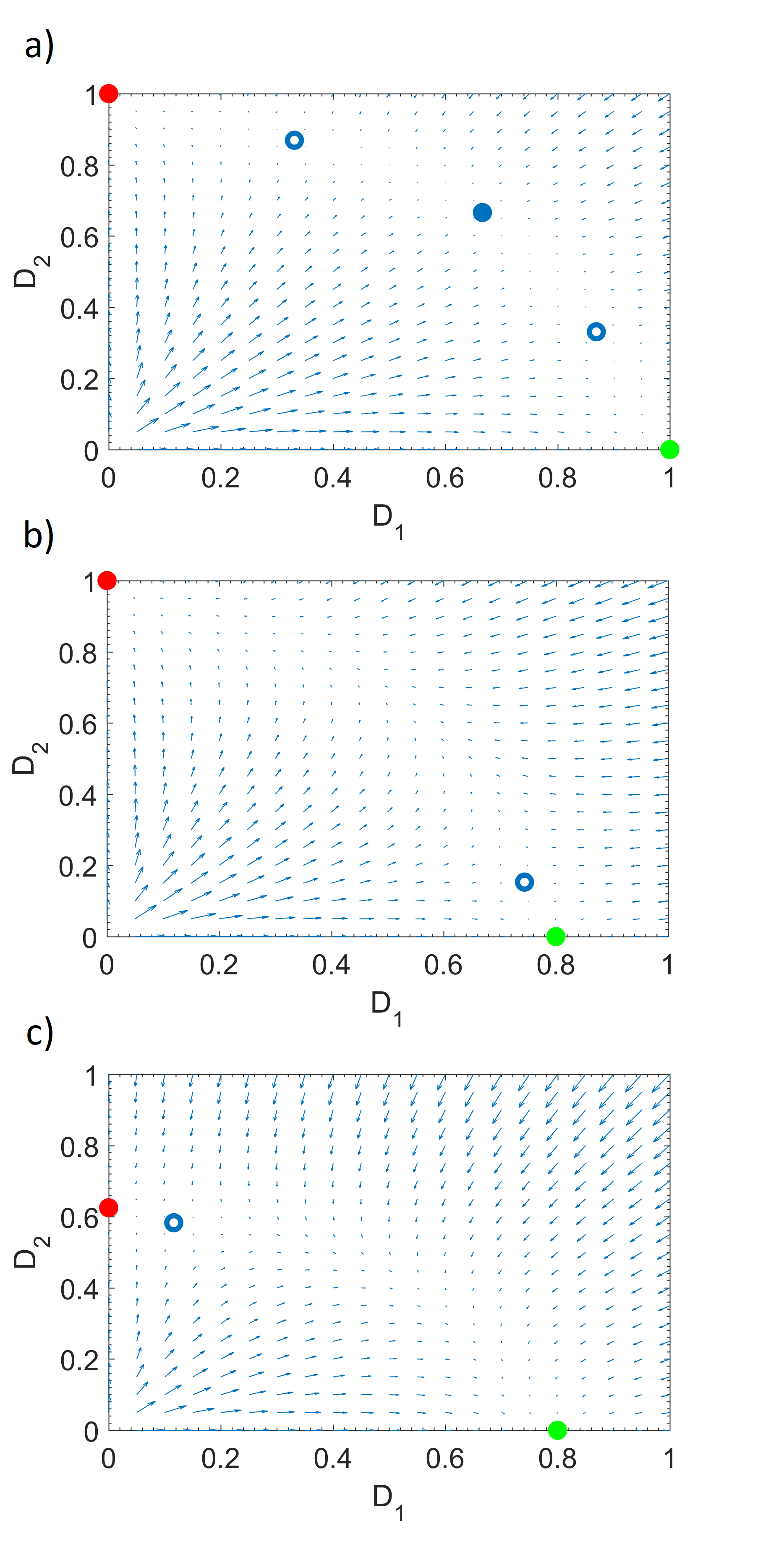}
		\caption{\label{fig:figure3}Fixed points $(D_1^*, D_2^*)$ of the secular equations. The arrows indicate the flow. Stable (unstable) fixed points are represented by solid (hollow) circles. The green (red) stable fixed point correspond to the system choosing the path with minimal (maximal) average dissipation. Subplot (a) corresponds to the case of equal, constant dissipation. In subplot (b) $\langle \Phi_1 \rangle = 0$ and $\langle \Phi_2 \rangle = 0.6$, while in (c) $\langle \Phi_1 \rangle = 0.25$ and $\langle \Phi_2 \rangle = 0.6$.}
	\end{figure}
In the following, we will perform a linear stability analysis of the system of differential Eq.~\eqref{Eq:D}. In the stationary regime, the conductivities $D_1$ and $D_2$ oscillate with the period $T = \frac{2 \pi}{\omega}$ of the dissipation around the stationary values $\langle D_1 \rangle = D_1^*$ and $\langle D_2 \rangle = D_2^*$ and it holds
\begin{align}
    \Big \langle \frac{\partial D_i}{\partial t} \Big \rangle & = \lim_{t \to \infty} \Big (\frac{1}{T} \int_{t}^{t + T} \frac{\partial D_i}{\partial t} (t') dt' \Big ) \nonumber
    \\ & = 0\,.
\end{align}
At this point, we remind the reader that we use $\tilde D_i = D_i/(D_1 + D_2)$. Applying this averaging procedure to Eq.~\eqref{Eq:D} we obtain:
\begin{align}
    0 = \langle f(\tilde D_i) \rangle - \langle \gamma_i(t) D_i \rangle\,.
\end{align}
We assume that the conductivities $D_1$ and $D_2$ are approximately constant during the period $T$. Thus, in the stationary regime the relation holds:
\begin{align}
    \langle \gamma_i(t) D_i \rangle & = \lim_{t \to \infty} \Big (\frac{1}{T} \int_{t}^{t + T} \gamma_i(t') D_i dt' \Big ) \nonumber
    \\ & \approx \lim_{t \to \infty} \Big (\frac{1}{T} D_i^* \int_{t}^{t + T} \gamma_i(t') dt' \Big ) = \gamma_i^{\text{eff}} D_i^*\,.
\end{align}
This leads us to the equation
\begin{align}
    \label{Eq:average}
    0 = \langle f(\tilde D_i) \rangle - \gamma_i^{\text{eff}} D_i^*\,.
\end{align}
Performing a Taylor expansion of the first term in this equation around the stationary values $D_1^*$ and $D_2^*$ to first order, we get
\begin{align}
    \langle f(\tilde D_i) \rangle \approx\,\,& f(\tilde D_i^*)\, \nonumber \\ & + \Big (\frac{\partial f}{\partial \tilde D_i} \frac{\partial \tilde D_i}{\partial D_1} \Big ) \Big |_{D_1 = D_1^*, D_2 = D_2^*} \langle (D_1 - D_1^*) \rangle \nonumber \\ & + \Big (\frac{\partial f}{\partial \tilde D_i} \frac{\partial \tilde D_1}{\partial D_2} \Big ) \Big |_{D_1 = D_1^*, D_2 = D_2^*} \langle (D_2 - D_2^*) \rangle.
\end{align}
We notice that the terms of first order vanish due to $\langle (D_i - D_i^*) \rangle = \langle D_i \rangle - D_i^* = 0$. Using this expression, Eq.~\eqref{Eq:average} yields
\begin{align}
    \label{Eq:fixed}
    0 = f(\tilde D_i^*) - \gamma_i^{\text{eff}} D_i^*\,.
\end{align}
In the following, we introduce the deviation $y_i = D_i - D_i^*$ of the conductivities from their stationary value. We can cast Eq.~\eqref{Eq:D} in the form
\begin{align}
    \label{Eq:y_i}
    \frac{\partial y_i}{\partial t} & = f(\tilde D_i) - \gamma_i(t) (y_i + D_i^*)\,.
\end{align}
Assuming that the period $T$ of the illumination is much smaller than the time scale on which the network structure changes, we can cast Eq.~\eqref{Eq:y_i} in the form
\begin{align}
    \frac{\partial y_i}{\partial t} & = f(\tilde D_i) - \gamma_i^{\text{eff}} (y_i + D_i^*)\,.
\end{align}
Performing a Taylor expansion of the first term in this equation around the stationary values $D_1^*$ and $D_2^*$ to first order, we get
\begin{align}
    \label{Eq:taylor}
    f(\tilde D_i) \approx & \,\, f(\tilde D_i^*) \nonumber \\ & + \Big (\frac{\partial f}{\partial \tilde D_i} \frac{\partial \tilde D_i}{\partial D_1} \Big ) \Big |_{D_1 = D_1^*, D_2 = D_2^*} y_1 \nonumber \\ & + \Big (\frac{\partial f}{\partial \tilde D_i} \frac{\partial \tilde D_1}{\partial D_2} \Big ) \Big |_{D_1 = D_1^*, D_2 = D_2^*} y_2 \nonumber
    \\ = & \,\, f(\tilde D_i^*) + A_i y_1 + B_i y_2\,,
\end{align}
with
\begin{align}
    \label{Eq:A_i}
    A_i = \Big (\frac{\partial f}{\partial \tilde D_i} \frac{\partial \tilde D_i}{\partial D_1} \Big ) \Big |_{D_1 = D_1^*, D_2 = D_2^*} \nonumber
    \\ B_i = \Big (\frac{\partial f}{\partial \tilde D_i} \frac{\partial \tilde D_1}{\partial D_2} \Big ) \Big |_{D_1 = D_1^*, D_2 = D_2^*}\,.
\end{align}
Combining Eq.~\eqref{Eq:y_i} and Eq.~\eqref{Eq:A_i} with Eq.~\eqref{Eq:taylor} yields
\begin{align}
    \label{Eq:system}
    \frac{\partial y_i}{\partial t} & = A_i y_1 + B_i y_2 - \gamma_i^{\text{eff}} (y_i + D_i^*) + \gamma_i^{\text{eff}} D_i^* \nonumber
    \\ & = A_i y_1 + B_i y_2 - \gamma_i^{\text{eff}} y_i\,.
\end{align}
In order to solve the system of differential Eq.~\eqref{Eq:system}, we make the ansatz $y_i(t) = Y_i e^{\lambda t}$ with $\lambda \in \mathbb C$. This yields
\begin{align}
    \underbrace{
    \begin{pmatrix}
    \lambda - A_1 + \gamma_1^{\text{eff}} & - B_1
    \\ - A_2 & \lambda - B_2 + \gamma_2^{\text{eff}}
    \end{pmatrix}
    }_{A :=}
    \begin{pmatrix}
    Y_1
    \\ Y_2
    \end{pmatrix}
    =
    \begin{pmatrix}
    0
    \\ 0
    \end{pmatrix}
\end{align}
To find non-trivial solutions, it must hold $\text{det}(A) = 0$ which gives
\begin{align}
    0 = & \,\, \text{det}(A) \nonumber
    \\ \Leftrightarrow 0 = & \,\,(\lambda - A_1 + \gamma_1^{\text{eff}}) (\lambda - B_2 + \gamma_2^{\text{eff}}) - A_2 B_1 \nonumber
    \\ \Leftrightarrow \lambda_{1, 2} = & \,\, \frac{A_1 + B_2 - (\gamma_1^{\text{eff}} + \gamma_2^{\text{eff}})}{2} \nonumber \\ & \pm \bigg (\frac{A_1 + B_2 - (\gamma_1^{\text{eff}} + \gamma_2^{\text{eff}})^2}{4}  \nonumber \\ & \text{\,\,\,\,\,\,\,\,\,\,} - (A_1 - \gamma_1^{\text{eff}})(B_2 - \gamma_2^{\text{eff}}) + A_2 B_1 \bigg )^{1/2}\,.
\end{align}
The fixed points $(D_1^*, D_2^*)$ are given by Eq.~\eqref{Eq:fixed}. They are stable if the corresponding values $\lambda_i$ are negative. The corresponding fixed points $c^*$ of this quantity can be calculated from the fixed points $(D_1^*, D_2^*)$. A fixed point $c^*$ is considered stable if $(D_1^*, D_2^*)$ is a stable fixed point. Figure \ref{fig:figure3} shows the fixed points for different choices of the average dissipation along the two paths. 

\section{Fokker-Planck equation}
\label{App:D}

The variable $\tilde{c}$ is assumed to undergo a time-continuous Markov process in the presence of a drift $\mu(\tilde c)$ and a Ito-diffusion with amplitude $\sigma(\tilde c,t)$:
\begin{align}
\label{Eq:Ito}
	    \frac{\partial \tilde c}{\partial t} = \mu(\tilde c) + \sigma(\tilde c) \xi(t)
\end{align}
with $\xi(t)$ describing white noise. Drift and diffusion coefficient are determined by means of equation-free analysis:
\begin{align}
	    \mu(\tilde c) & = \frac{\langle c(t + \delta t) - c(t) | c(t) = \tilde c \rangle_E}{\delta t}\,,\\ 
	    \sigma^2(\tilde c) & = \frac{\langle (c(t + \delta t) - c(t) - \mu(\tilde c) \delta t)^2 | c(t) = \tilde c \rangle_E}{\delta t}\,,
\end{align}
where $\langle \cdot\rangle_E$ indicates a sample average and $c(t)$ is calculated from Eq.~\eqref{Eq:c} from the values of $D_i(t)$, that are obtained by numerical integration of Eq.~\eqref{Eq:D}. We verify the validity of this approach by comparing the values of $c_{\rm end}$ we obtain by numerically integrating Eq.~\eqref{Eq:Ito} with the ones of Eq.~\eqref{Eq:D}, see Fig.\ref{fig:1D-2D}(a). The Fokker-Planck equation corresponding to Eq.~\eqref{Eq:Ito} is given in Eq.~\eqref{Eq:FPE}.

\end{appendix}


\begin{thebibliography}{99}

\bibitem{Cross:1993} 
M.\ C.\ Cross and P.\ C.\ Hohenberg, Pattern formation outside of equilibrium, Rev. Mod. Phys. {\bf 65}, 851 (1993).

\bibitem{Sherrington:2010}
D.\ Sherrington, Physics and complexity, Phil. Trans. R. Soc. A {\bf 368}, 1175 (2010).

\bibitem{Zia:2011}
 T.\ Chou, K.\ Mallick, and R.\ K.\ P.\ Zia, "Non-equilibrium statistical mechanics: From a paradigmatic model to biological transport",
 Reports on Progress in Physics {\bf 74}, 116601, (2011).
 
 \bibitem{Borgani:2012}
A.\ V.\ Kravtsov and S.\ Borgani, Formation of Galaxy Clusters, Annual Review of Astronomy and Astrophysics {\bf 50}, 353 (2012).
 
 \bibitem{KPZ:1986}
 M.\ Kardar, G.\ Parisi, and Y.-C.\ Zhang, Dynamic Scaling of Growing Interfaces, Phys. Rev. Lett. {\bf 56} (1986).
 
 \bibitem{Meron:2001}
 J.\ von Hardenberg, E.\ Meron, M.\ Shachak, and Y.\ Zarmi, Diversity of vegetation patterns and desertification, Phys. Rev. Lett. {\bf 87}, 198101 (2001).
 
 \bibitem{Steinbock:2019}
A.-K.\ Malchow, A.\ Azhand, P.\ Knoll, H.\ Engel, and O.\ Steinbock, From nonlinear reaction-diffusion processes to permanent microscale structures, Chaos: An Interdisciplinary Journal of Nonlinear Science, {\bf 29}, 053129 (2019).

\bibitem{Folz:2019}
F.\ Folz, L.\ Wettmann, G.\ Morigi, and K.\ Kruse, Sound of an axon's growth, Phys. Rev. E {\bf 99}, 050401(R) (2019).

\bibitem{Tero:2007}
A.\ Tero, R.\ Kobayashi, and T.\ Nakagaki, A mathematical model for adaptive transport network in path finding by true slime mold, J Theor Biol. {\bf 244}, 553-64 (2007).

\bibitem{Nakagaki:2000}
T. Nakagaki, H. Yamada and A. Tero, Maze-solving by an amoeboid organism, Nature {\bf 407}, 470 (2000).

\bibitem{Nakagaki:2004}
T. Nakagaki, H. Yamada and M. Hara, Smart network solutions in an amoeboid organism, Biophysical Chemistry, {\bf 107} (2004).

\bibitem{Oettmeier:2020}
C.\ Oettmeier, T.\ Nakagaki, and H.\ G.\ D\"obereiner, Slime mold on the rise: the physics of Physarum polycephalum, Journal of Physics D: Applied Physics {\bf 53}, 310201 (2020).

\bibitem{Tero:2010}
A.\ Tero, S.\ Takagi, T.\ Saigusa, K.\ Ito, D.\ P.\ Bebber, M.\ D.\ Fricker, K.\ Yumiki, R.\ Kobayashi, and T.\ Nakagaki, Rules for Biologically Inspired Adaptive Network Design, Science {\bf 327}, 5964 (2010)

\bibitem{Boussard:2021}
A.\ Boussard, A.\ Fessel, C.\ Oettmeier, L.\ Briard, H.-G.\ D\"obereiner,
and A.\ Dussutour, Adaptive behaviour and learning in slime moulds: the role of oscillations, Philosophical Transactions of the Royal Society B {\bf 376}, 20190757 (2021).
  
\bibitem{Gao:2019} 
Chao Gao, Chen Liu, D.\ Schenz, Xuelong Li, Zili Zhang, M.\ Jusup, Zhen Wang, M.\ Beekman, and T.\ Nakagaki,  Does being multi-headed make you better at solving problems? A survey of Physarum-based models and computations, Physics of Life Reviews {\bf 29}, 1 (2019); Physarum inspires research beyond biomimetic algorithms, {\it ibidem} {\bf 29}, 51 (2019).

\bibitem{Gross:2005}
C.\ Gross, Complex and Adaptive Dynamical Systems (Springer, Berlin, 2013)

\bibitem{Flies:2018}
J.\ Noetel, V.\ L.\ S.\ Freitas, E.\ E.\ N.\ Macau, and L.\ Schimansky-Geier, Optimal noise in a stochastic model for local search, Phys. Rev. E {\bf 98}, 022128 (2018).

\bibitem{Meyer:2017}
B.\ Meyer, C.\ Ansorge, and T.\ Nakagaki, The role of noise in self-organized decision making by the true slime mold Physarum polycephalum, PLoS ONE {\bf 12}, e0172933 (2017).

\bibitem{Bonifaci:2020}
V.\ Bonifaci, E.\ Facca, F.\ Folz, A.\ Karrenbauer, P.\ Kolev, K.\ Mehlhorn, G.\ Morigi, G.\ Shahkarami, and Q.\ Vermande, Physarum Multi-Commodity Flow Dynamics, preprint arxiv:2009.01498 (2020)

\bibitem{Haken:1983}
H.\ Haken, Synergetics, an Introduction: Nonequilibrium Phase Transitions and Self-Organization in Physics, Chemistry, and Biology (Springer-Verlag, New York, 1983).

\bibitem{vanKampen}
N.\ G.\ Van Kampen, Stochastic Processes in Physics and Chemistry (Elsevier, Amsterdam, 2007).

\bibitem{Kloeden}
P.\ E.\ Kloeden and E.\ Platen, Numerical Solution of Stochastic Differential Equations (Springer, Berlin, 1992).

\bibitem{Gammaitoni:1998}
L.\ Gammaitoni, P.\ H\"anggi, P. \ Jung, and F.\ Marchesoni, Stochastic Resonance, Rev. Mod. Phys. {\bf 70}, 225 (1998). 

\bibitem{Meyer:2017b}
B.\ Meyer, Optimal information transfer and stochastic resonance in collective decision making, Swarm Intell. {\bf 11}, 131 (2017).

%\bibitem{Gupta:2013}
%S. Gupta, A. Campa, and S. Ruffo, Kuramoto model of synchronization: equilibrium and nonequilibrium aspects, J. Stat. Mech.: Theory Exp. R08001 (2014).


\end{thebibliography}
\end{document}